\documentstyle{article}
\begin{document}
\newcommand{\beq}{\begin{equation}}
\newcommand{\eeq}{\end{equation}}
\newcommand{\bea}{\begin{eqnarray}}
\newcommand{\eea}{\end{eqnarray}}
\newcommand{\half}{\frac{1}{2}}
\newcommand{\ihalf}{\frac{i}{2}}
\newcommand{\param}{\epsilon}
\newcommand{\opar}{\overline{\eta}}
\newcommand{\ghalf}{\frac{g}{2}}
\newcommand{\quart}{\frac{1}{4}}
\newcommand{\cA}{\cal A}
\newcommand{\cB}{\cal B}
\newcommand{\cH}{\cal H}
\newcommand{\suma}{\sum_{q=1}^2}
\newcommand{\dmu}{\partial_{\mu}}
\newcommand{\dmup}{\partial^{\mu}}
\newcommand{\aslash}{\not\!\!\cA}
\newcommand{\hslash}{\not\!\!\cH}
\newcommand{\Dslash}{\not\!\! D}
\newcommand{\dslash}{\not\!\partial}
\renewcommand{\O}{\Omega}
\renewcommand{\a}{\alpha}
\renewcommand{\t}{\theta}
\renewcommand{\S}{\Sigma}
\newcommand{\g}{\gamma}
\renewcommand{\l}{\lambda}
\newcommand{\luq}{\lambda_1^{(q)}}
\newcommand{\ldq}{\lambda_2^{(q)}}
\newcommand{\pq}{\Phi_{(q)}}
\newcommand{\pqd}{\Phi_{(q)}^{\dag}}
\newcommand{\opq}{\overline{\Psi}_{(q)}}
\newcommand{\psiq}{\Psi_{(q)}}
\newcommand{\duum}{\partial^{\mu}N}
\newcommand{\mps}{\vert\phi\vert^2}
\newcommand{\oSig}{\overline{\Sigma}}
\newcommand{\ol}{\overline{\l}}
\newcommand{\pau}{\vec{\tau}}
\newcommand{\fiq}{\Phi_{(q)}}
\newcommand{\opsi}{\overline{\psi}}
\newcommand{\oW}{\overline{W}}
\newcommand{\oD}{\overline{D}}
\newcommand{\tsq}{\theta^2}
\newcommand{\otsq}{\overline{\theta}^2}
\newcommand{\ts}{\overline{\theta}{\theta}}
\newcommand{\tu}{\t\sigma^{\mu}\tt}
\renewcommand{\tt}{\overline{\theta}}
\newcommand{\G}{\Gamma}

\title
{A Topological Bound for Electroweak 
Vortices from Supersymmetry}
\author{
Jos\'e D. Edelstein\thanks{Consejo Nacional de Investigaciones 
Cient\'{\i}ficas y T\'ecnicas.} \hspace{.7pt} and
Carlos N\'u\~nez \\
Departamento de F\'\i sica, Universidad Nacional de La Plata\\
C.C. 67, (1900) La Plata\\
Argentina\\ 
{\small hep-th/9507102}}

\date{}
\maketitle

\begin{abstract}
We study the connection between $N=2$ supersymmetry and 
a topological bound in a two-Higgs-doublet system having an
$SU(2)\times U(1)_Y\times U(1)_{Y^{\prime}}$ gauge group.
We derive Bogomol'nyi equations from supersymmetry considerations
showing that they hold provided certain conditions
on the coupling constants, which are a consequence of the huge 
symmetry of the theory, are satisfied.
\end{abstract}

\pagenumbering{arabic}

Supersymmetric Grand Unified Theories (SUSY GUTs) have
attracted much attention in connection with the hierarchy problem 
in possible unified theories of strong and electroweak
interactions \cite{R,PDB}.
In view of the requirement of electroweak symmetry breaking, 
these models necessitate an enrichment of the Higgs sector \cite{NACh},
posing many interesting questions both from the classical and quantum 
point of view. In fact, many authors 
have explored the existence of stable vortex 
solutions in a variety of multi-Higgs systems \cite{DSPEJL,BL} which
mimic the bosonic sector of SUSY GUTs, as it happens in the abelian
Higgs model \cite{NO}.

Vortices emerging as finite energy solutions of gauge 
theories can be usually shown to satisfy a topological bound for 
the energy, the so-called Bogomol'nyi bound \cite{BdVS}. 
Bogomol'nyi bounds were shown to reflect the presence of an extended 
supersymmetric structure \cite{WO}-\cite{ENS} - this requiring certain 
conditions on coupling constants - where the central charge coincides
with the topological charge. Being originated in the supercharge 
algebra, the bound is expected to be exact quantum mechanically.

Since multi-Higgs models can be understood to be motivated by SUSY GUTs, 
Supersymmetry is a natural framework to investigate Bogomol'nyi bounds. 
We shall study, then, the supersymmetric generalization of the 
$SU(2)\times U(1)_Y\times U(1)_{Y^{\prime}}$ model with two-Higgs 
first introduced in Ref.\cite{BL}. The theory has the same gauge group 
structure as that of supersymmetric extensions of the 
Weinberg-Salam Model that arise
as low energy limits of $E_6$ based Grand Unified or superstring 
theories.
In spite of being a simplified model (in the sense that its Higgs
structure is not so rich as that of Grand Unified theories), 
it can be seen as the minimal extension of the Standard Model 
necessary for having Bogomol'nyi equations. 
We show that the Bogomol'nyi bound of the model, as well as 
the Bogomol'nyi equations, are straight consequences of the 
requirement of $N=2$ supersymmetry imposed on the theory. 
We also show explicitely that a necessary condition to achieve 
the $N=2$ model implies certain relations between coupling constants
that equal those found in \cite{BL} for the existence of a Bogomol'nyi
bound. 

The $SU(2)\times U(1)_Y\times U(1)_{Y^{\prime}}$ gauge theory in
$2+1$, introduced in Ref.\cite{BL}, is described by the action
\beq
S = \int d^3x \left[ - \quart \vec{W}_{\mu\nu}\cdot\vec{W}^{\mu\nu}
- \quart F_{\mu\nu}F^{\mu\nu} - \quart G_{\mu\nu}G^{\mu\nu} 
+ \half\suma\vert{\cal D}_{\mu}^{(q)}\Phi_{(q)}\vert^2 
- {V}(\Phi_{(1)},\Phi_{(2)}) \right] 
\label{acbos}
\eeq
where $\Phi_{(1)}$ and $\Phi_{(2)}$ are a couple of 
Higgs doublets under the $SU(2)$ factor of the gauge group,
$A$ and $B$ are real scalar fields and $\vec{W} = W^a\tau^a$ is a real
scalar in the adjoint representation of $SU(2)$.
The specific form of the potential will be determined below.
The strength fields can be written in terms of the gauge fields 
$A_{\mu}$, $B_{\mu}$ and $\vec{W}_{\mu}$. The covariant derivative 
is defined as:
\beq
{\cal D}_{\mu}^{(q)}\fiq = \left( \partial_{\mu} + 
\frac{i}{2}gW_{\mu}^a\tau^a + \frac{i}{2}\alpha_{(q)}A_{\mu} 
+ \frac{i}{2}\beta_{(q)}B_{\mu} \right) \fiq, 
\;\;\;\;\; \mbox{\scriptsize q=1,2}
\label{dercov}
\eeq
where $g$ is the $SU(2)$ coupling constant 
while $\alpha_{(q)}$ and $\beta_{(q)}$ represents the different 
couplings of $\Phi_{(q)}$ with $A_{\mu}$ and $B_{\mu}$.
A minimal $N=1$ supersymmetric extension of this model is given by an
action which in superspace reads:
\bea
{\cal S}_{N=1} & = &  \half \int d^3x d^2\theta
\left[{\overline{\Omega}}_A\Omega_A + {\overline{\Omega}}_B\Omega_B 
+ {\overline{\Omega}}^a_{\vec{W}}{\Omega}^a_{\vec{W}} 
- \overline{{\cal D}{\cal A}}{\cal D}{\cal A} - \overline{{\cal 
D}{\cal B}}{\cal D}{\cal B} - \overline{{\cal D}{\cal W}}^a{\cal 
D}{\cal W}^a + \xi_1{\cal A} + \xi_2{\cal B} \right.\nonumber\\
& + & \left. \half\sum_{q=1}^2
\left[(\overline{\nabla^{(q)}\Upsilon_{(q)}})^a
({\nabla^{(q)}\Upsilon_{(q)}})^a + 
i\Upsilon_{(q)}^{\dag}\left(\sqrt{2\luq}{\cal A} + \sqrt{2\ldq}{\cal 
B} + \sqrt{2\l_3}{\cal W}^a\tau^a\right)\Upsilon_{(q)}\right] 
\right]
\label{supaction}
\eea
where
\beq
{\nabla^{(q)}\Upsilon_{(q)}}= \left({\cal D} + \ihalf 
g[\Gamma_{\vec{W}},] - \ihalf \alpha_{(q)}[\Gamma_A,] 
+ \ihalf \beta_{(q)}[\Gamma_B,]\right)
\Upsilon_{(q)}.
\label{nablaq}
\eeq
$\Upsilon_{(q)} \equiv (\Phi_{(q)},\Psi_{(q)})$ are a couple 
of complex doublet superfields, ${\cal A} \equiv (A,\chi_A)$, 
${\cal B} \equiv (B,\chi_B)$ and ${\cal W} \equiv 
(W^a,\chi_{\vec{W}}^a)\tau^a$ are real scalar superfields and 
$\Gamma_A \equiv (A_{\mu},\rho_A)$, $\Gamma_B \equiv
(B_{\mu},\rho_B)$ and $\Gamma_{\vec{W}} \equiv 
\Gamma_{\vec{W}}^a\tau^a = (W_{\mu}^a,\lambda^a)\tau^a$ are three 
spinor gauge superfields in the Wess-Zumino gauge.
$\Omega_A$, $\Omega_B$ and ${\Omega}^a_{\vec{W}}$, are the 
corresponding superfield strengths. Concerning $\lambda_1^{(q)}$,
$\lambda_2^{(q)}$, $\lambda_3$, $\xi_1$ and $\xi_2$, they are real
constants whose significance will be clear below.
Finally, ${\cal D}$ is the usual supercovariant derivative, 
${\cal D} = \partial_{\overline{\theta}} + 
i\overline{\theta}\dslash$, while the $\gamma$-matrices are represented 
by $\gamma^0 = \tau^3$, $\gamma^1 = i\tau^1$ and $\gamma^2 = 
-i\tau^2$. In the sake of simplicity, we shall consider configurations 
with vanishing $A$,
$B$ and $\vec{W}$\footnote{The 
$SU(2)\times U(1)_Y\times U(1)_{Y^{\prime}}$ theory with its full 
field content is considered in Ref.\cite{EN}.}. 
Then, the Higgs potential in (\ref{supaction}) 
takes the form:
\beq
{V}(\Phi_{(1)},\Phi_{(2)}) = \left(\suma\sqrt{\luq}\pqd\pq 
- \frac{\xi_1}{\sqrt{2}}\right)^2 + \left(\suma\sqrt{\ldq}\pqd\pq 
- \frac{\xi_2}{\sqrt{2}}\right)^2 + 
\l_3\left(\suma\pqd\tau^a\pq\right)^2. 
\label{invpot}
\eeq
In order to extend the supersymmetric invariance of the 
theory to $N=2$, we consider transformations with a complex parameter
\cite{LLW,ENS}. We first combine all the spinors into Dirac 
fermions as:
\beq
\Sigma_A \equiv \chi_A - i\rho_A \;\;\; , \;\;\;
\Sigma_B \equiv \chi_B - i\rho_B \;\;\; \mbox{and} \;\;\;
\Xi^a \equiv \chi_{\vec{W}}^a - i\Lambda^a.
\label{sigmas}
\eeq
The fermionic contribution to the (non-minimal part of the) 
interaction lagrangian can be written as
\bea
{L}_{Fer,int} & = & \suma\left[\frac{\alpha_{(q)}+\sqrt{8\luq}}{4}
\opq\Sigma_A\pq + \frac{\beta_{(q)}+\sqrt{8\ldq}}{4} \opq\Sigma_B\pq 
+ \frac{g+\sqrt{8\l_3}}{4} \opq\Xi^a\tau^a\pq \right. \nonumber \\
& - & \left.\frac{\alpha_{(q)}-\sqrt{8\luq}}{4} \opq\tilde\Sigma_A\pq 
- \frac{\beta_{(q)}-\sqrt{8\ldq}}{4} \opq\tilde\Sigma_B\pq - 
\frac{g-\sqrt{8\l_3}}{4} \opq\tilde\Xi^a\tau^a\pq \right],
\label{acferult}
\eea
where $\tilde\Xi^a$, $\tilde\Sigma_A$ and $\tilde\Sigma_B$ are the 
charge conjugates of $\Xi^a$, $\Sigma_A$ and $\Sigma_B$ respectively.
Now, transformations with complex parameter $\eta$ 
are equivalent to transformations with a real parameter
followed by a phase transformation for fermions,  
$\{\Xi^a,\Sigma_A,\Sigma_B,\Psi_{(q)}\} \longrightarrow 
e^{i\alpha}\{\Xi^a,\Sigma_A,\Sigma_B,\Psi_{(q)}\}$.
Then, $N=2$ supersymmetry requires invariance under this fermion
rotation.
One can easily see from (\ref{acferult})
that fermion phase rotation invariance is achieved
if and only if:
\beq
\l_3 = \frac{g^2}{8} \;\;\;\;\;\;\; , \;\;\;\;\;\;\;
\luq = \frac{\alpha_{(q)}^2}{8} \;\;\;\;\;\;\; \mbox{and} 
\;\;\;\;\;\;\;
\ldq = \frac{\beta_{(q)}^2}{8}.
\label{conds}
\eeq
These conditions, imposed by the requirement of extended supersymmetry, 
fix the coupling constants exactly as they appear in \cite{BL}.
Thus, we have shown that the potential and the coupling constants of the
$SU(2)\times U(1)_Y\times U(1)_{Y^{\prime}}$ model are simply dictated 
by $N=2$ supersymmetry. This result is analogous to that recently 
found in the Abelian Higgs model \cite{ENS}. 

We shall now analyse the $N=2$ algebra of supercharges for our model. 
To construct the conserved charges we follow the Noether method and
obtain ${\cal Q}[\eta] \equiv \overline{\eta}Q + \overline{Q}\eta$,
with
\bea
\overline{Q} & = & - \ihalf\int d^2x
\left\{{\Sigma}_A^{\dag}\left[\half\epsilon^{\mu\nu\lambda}F_{\mu\nu}
\gamma_{\lambda} + \sum_{q=1}^{2}\frac{\alpha_{(q)}}{2}\pqd\pq 
- \xi_1 \right] + 
{\Sigma}_B^{\dag}\left[\half\epsilon^{\mu\nu\lambda}G_{\mu\nu}
\gamma_{\lambda} + \sum_{q=1}^{2}\frac{\beta_{(q)}}{2}\pqd\pq
\right.\right.\nonumber\\
& - & \left.\left. \xi_2 \right]
+ {\Xi}^{\dag a}\left[\half\epsilon^{\mu\nu\lambda}W_{\mu\nu}^a
\gamma_{\lambda} + \frac{g}{2}\sum_{q=1}^{2}\pqd\tau^a\pq 
\right] - i\sum_{q=1}^{2}\psiq^{\dag}\gamma^{\mu}{\cal 
D}_{\mu}^{(q)}\pq \right\}
\label{Q}
\eea
Since we are interested in connecting the $N=2$
supercharge algebra with Bogomol'nyi
equations and bound, 
we impose static configurations with $A_0 = B_0 = W_0^a = 0$, 
and we restrict ourselves to the bosonic sector of the theory after
computing the algebra. We obtain, after some calculations
\beq
\{\bar{\cal Q},{\cal Q}\} = 2\opar\gamma_{0}\eta P^{0}
+ \opar\eta Z
\label{qeqe}
\eeq
where
\beq
P^0 = E = \half \int d^2x \left[ \half (W_{ij}^a)^2 + 
\half (F_{ij})^2 + \half G^2_{ij} + 
\suma |{\cal D}_i^{(q)}\Phi_{(q)}|^2 + V(\Phi_{(1)},\Phi_{(2)})\right]
\label{po}
\eeq
while the central charge is given by:
\bea
Z & = & - \int d^2x \left[ \half\epsilon^{ij}F_{ij}\left(
\suma\frac{\alpha_{(q)}}{2}\Phi_{(q)}^{\dag}\Phi_{(q)} - \xi_1\right) 
+ \half\epsilon^{ij}G_{ij}\left(
\suma\frac{\beta_{(q)}}{2}\Phi_{(q)}^{\dag}\Phi_{(q)} 
- \xi_2\right) \right.\nonumber\\
& + & \left. 
\frac{g}{4}\epsilon^{ij}W^a_{ij}\suma\Phi_{(q)}^{\dag}\Phi_{(q)} 
+ i\epsilon^{ij}\suma ({\cal D}^{(q)}_i\Phi_{(q)})({\cal 
D}^{(q)}_j\pq)^* \right] = \half \oint {\cal V}_i dx^i,
\label{zeta}
\eea
where ${\cal V}^i$ is given by
\beq
{\cal V}^i =  \left(\xi_1A_j + \xi_2B_j + 
i\suma \Phi_{(q)}^{\dag}{\cal D}^{(q)}_j\pq \right)\epsilon^{ij}.
\label{vi}
\eeq
Finite energy dictates the following asymptotic behaviour for the 
Higgs doublets \cite{BL}
\beq
{\Phi_{(1)}}_{\infty} = \frac{\phi_0}{\sqrt{2}}
\left( \begin{array}{c} 0 \\ \exp{in_{(1)}\varphi} \end{array} \right)
\;\;\; , \;\;\;
{\Phi_{(2)}}_{\infty} = \frac{\phi_0}{\sqrt{2}}
\left( \begin{array}{c} \exp{in_{(2)}\varphi} \\ 0 \end{array} \right),
\label{asymp3}
\eeq
where $n_{(1)}$ and $n_{(2)}$ are integers that sum up to the 
topological charge of the configuration $m$,
\beq
m \equiv n_{(1)} + n_{(2)}. 
\label{top}
\eeq
Then, coming back to eq.(\ref{zeta}) for the central charge, 
after Stokes' theorem, we see that
\beq
Z = \oint (\xi_1A_i + \xi_2B_i)dx^i = - 4\pi\phi_0^2 m
\label{zetaint}
\eeq
that is, the central charge of the $N=2$ algebra equals the 
topological charge of the configuration.
It is now easy to find the Bogomol'nyi bound from the
supersymmetry algebra (\ref{qeqe}). Indeed,
\beq
\{\bar{\cal Q},{\cal Q}\} = \int d^2x \left[
(\delta\Xi^a)^{\dag}(\delta\Xi^a) + 
(\delta\Sigma_A)^{\dag}(\delta\Sigma_A) + 
(\delta\Sigma_B)^{\dag}(\delta\Sigma_B) 
+ \suma (\delta \Psi_{(q)})^{\dag}(\delta \Psi_{(q)})\right] 
\geq 0.
\label{cota}
\eeq
the lower bound being saturated if and only if $\delta\Xi^a 
= \delta\Sigma_A = \delta\Sigma_B = \delta 
\Psi_{(q)} = 0$. Non-trivial solutions to these equations force us to 
choose a parameter with definite chirality, say $\eta_+$. Now,
conditions
\beq 
\delta_{\eta_+}\Xi^a = 
\delta_{\eta_+}\Sigma_A = 
\delta_{\eta_+}\Sigma_B = 
\delta_{\eta_+}\Psi_{(q)} = 0
\label{etamas}
\eeq
are nothing but the Bogomol'nyi equations of the theory:
\beq 
\half\epsilon^{ij}F_{ij} + 
\sum_{q=1}^{2}\frac{\alpha_{(q)}}{2}\pqd\pq - \xi_1 = 0 \;\;\;\;\; ,
\;\;\;\;\; \half\epsilon^{ij}G_{ij}
+ \sum_{q=1}^{2}\frac{\beta_{(q)}}{2}\pqd\pq - \xi_2 = 0 
\label{bogo1}
\eeq
\beq 
\epsilon^{ij}{W_{ij}}^a + g\sum_{q=1}^{2}\pqd\tau^a\pq  = 0 
\;\;\;\;\; \mbox{and} \;\;\;\;\; ({\cal D}_{i}^{(q)} - 
i\epsilon_{ij}{\cal D}_{j}^{(q)})\pq = 0.
\label{bogo2}
\eeq
Note that, for this chiral parameter, eq.(\ref{cota}) implies
the Bogomol'nyi bound of our model,
\beq
M \geq 2\pi\phi_0^2 m.
\label{bogobound}
\eeq

Let us remark on the fact that field configurations solving
Bogomol'nyi equations break half of the supersymmetries (those
generated by $\eta_-$), a common feature in all models presenting
Bogomol'nyi bounds with supersymmetric extension \cite{hull}. 
Were we faced with an antichiral parameter, we
would have obtained antisoliton solutions with broken of the
supersymmetry transformation generated by $\eta_+$.

The connection of our model with realistic supersymmetric extensions 
of the Standard model, 
and its coupling with supergravity (the possible
existence of string-like solutions in this last theory) remain open 
problems. We hope to report on these issues in a forthcoming work.

\end{document}